\documentclass[prd,twocolumn,superscriptaddress,amsfonts,amssymb,amsmath,showpacs]{revtex4-2}
\usepackage{bm}
\usepackage{mathtools}
\usepackage{lipsum}
\usepackage{latexsym}
\usepackage[latin1]{inputenc}
\usepackage{graphicx}
\usepackage{amsmath}
\usepackage{amssymb}
\usepackage{amsthm}
\usepackage{relsize}
\usepackage{palatino}
\usepackage{mathpazo}
\usepackage{textcomp}
\usepackage{float}
\usepackage{booktabs}
\usepackage{dcolumn}
\usepackage{booktabs}
\usepackage{multirow}
\usepackage{tikz}
\usepackage{hyperref}
\hypersetup{
    colorlinks=true,
    linkcolor=purple,
    filecolor=magenta,      
    citecolor=blue
}
\usepackage{amsmath}
\usepackage{xcolor}
\usepackage{orcidlink}
\usepackage[caption=true]{subfig}
\usepackage{commath}
\makeatletter
\long\def\dddddot#1{%
  {\mathop {#1}\limits ^{\vbox to-1.4\ex@ {\kern -\tw@ \ex@ \hbox {\normalfont .....}\vss }}}%
}
\long\def\multidots#1#2{%
  \count@=0
  {{\mathop {#2}\limits ^{\vbox to-1.4\ex@ {\kern -\tw@ \ex@ \hbox {\normalfont %
  \loop%
  \ifnum#1>\count@%
  .%
  \advance\count@ by1%
  \repeat%
  }\vss }}}}%
}
\makeatother

\begin{document}

\title{\bf Logarithmic and Strong Coupling Models in Weyl-Type $f(Q,T)$ Gravity}

\author{Rahul Bhagat\orcidlink{0009-0001-9783-9317}}
\email{rahulbhagat0994@gmail.com}
\affiliation{Department of Mathematics,
Birla Institute of Technology and Science-Pilani, Hyderabad Campus, Jawahar Nagar, Kapra Mandal, Medchal District, Telangana 500078, India.}

\author{S. K. Tripathy\orcidlink{0000-0001-5154-2297}}
\email{tripathy_sunil@rediffmail.com}
\affiliation{Department of Physics, Indira Gandhi Institute of Technology, Sarang, Dhenkanal, Odisha-759146, India.}

\author{B. Mishra\orcidlink{0000-0001-5527-3565}}
 \email{bivu@hyderabad.bits-pilani.ac.in}
 \affiliation{Department of Mathematics,
Birla Institute of Technology and Science-Pilani, Hyderabad Campus, Jawahar Nagar, Kapra Mandal, Medchal District, Telangana 500078, India.}

\begin{abstract}
In this paper, we have explored the cosmological implications of Weyl-type $f(Q,T)$ gravity, a modified gravitational theory formulated from Weyl geometry. The nonmetricity scalar $Q$ is coupled to the trace $T$ of the energy-momentum tensor. We analyze two models based on the logarithmic and strong coupling form of the function $f(Q,T)$. The corresponding field equations are then solved numerically after reformulating the system in terms of redshift. We employed the joint datasets CC+Union3.0+DESI DR2 and CC+Pantheon$^+$+DESI DR2 to perform a Markov Chain Monte Carlo (MCMC) analysis for constraining the model parameters. Using the constrained parameters, the geometrical and dynamical aspects of the models are analyzed. The results successfully describe a transition from decelerated to accelerated expansion for both models.  Both models exhibit quintessence-like behavior and approach the $\Lambda$CDM scenario at the present epoch ($z_0 = 0$). The Logarithmic model closely follows $\Lambda$CDM, approaching $q = -1$ at late times, while the Strong Coupling model shows similar deceleration for $0 < z < 1$ but transitions to phantom-like behavior ($z < 0$), suggesting a super-accelerated future expansion. The calculated age of the Universe from each model aligns with constraints from Planck and stellar age data. The violation of the strong energy condition and the satisfaction of null energy condition and dominant energy condition are shown.

\end{abstract}

\maketitle

\section{Introduction} 

The late-time cosmic phenomena has been one of the most transformative discoveries in modern cosmology. The evidence from distant Type Ia supernovae \cite{Riess_1998_116,Perlmutter_1998_517} and Cosmic Microwave Background (CMB) \cite{Komatsu_2011_192,Larson_2011_192} contradicts earlier expectations that cosmic expansion should slow down due to gravity\cite{Gibbons_1987,Hayashi_1979_19,Misner_1973_book}. The observed acceleration has been attributed to a mysterious component known as dark energy, which occupies a significant portion of the total energy budget of the Universe \cite{Nojiri_2007_04}. To begin with, the standard cosmological model, $\Lambda$CDM model is considered as the framework by introducing a cosmological constant $\Lambda$ as the simplest form of dark energy\cite{Riess_2019_876,Najera_2021_34,Junior_2016_33}. Though $\Lambda$CDM has been successful with large-scale structure of the Universe \cite{Raichoor_2020_500} and anisotropies in the CMB\cite{Balkenhol_2023_108}, however it suffers from theoretical issues. These include the cosmological constant problem, where quantum field theory predicts a vacuum energy vastly greater than what is observed \cite{Weinberg_1989_61}, and the coincidence problem, which questions the near-equality of matter and dark energy densities in the current epoch\cite{Cai_2005_2005_002,SULTANA2025101843}. Discrepancies in the measured value of the Hubble constant $H_0$ between early and late-Universe observations have led to significant disagreement, suggesting potential flaws in the $\Lambda$CDM paradigm \cite{DiValentino_2021_131}.\\

Since general relativity (GR) has limitations in addressing the recent behavior of the Universe, several modified theories of gravity have been formulated to explain late-time acceleration without invoking a cosmological constant\cite{Sotiriou_2010_82_451,Cai_2016_79,Sultana500354,Lokesh_Kumar_and_Yadav,SULTANA2023169392,SINGH2025117061}. In the nonmetricity approach, the gravitational theory is known as  
symmetric teleparallel gravity, in which the gravitational interaction is mediated by the non-metricity $Q$ \cite{Nester_1999_37_113}. This has been further refined into coincident GR or $f(Q)$ gravity \cite{Jimenez_2018_98}. Heisenberg \cite{Heisenberg_2019_796} offers a comprehensive and pedagogical account of metric-affine geometry, providing the mathematical foundation necessary for exploring the geometric trinity of gravity and its various generalizations. In the specific case of $f(Q)$ gravity, Chakraborty et al.\cite{Chakraborty_2025} investigated the reconstruction of $\Lambda$CDM like cosmological behavior by focusing on the three permissible symmetric teleparallel connections that are compatible with spatial homogeneity, isotropy and flatness. In a related study, Vasquez et al. \cite{R_Vasquez_2025} examined the cosmological dynamics within a flat FLRW background using an exponential two-parameter model of $f(Q)$ gravity. Their analytical approach presented a perturbative deviation from $\Lambda$CDM, enabling smooth evolution scenarios within the modified symmetric teleparallel gravity framework. Some of the prominent work on this gravity can be seen in Ref \cite{Barros_2020_30,SULTANA2025100422,Dimakis_2021_38,YADAV2024114}.  By introducing a non-minimal coupling between the nonmetricity $Q$ and the trace $T$ of the energy-momentum tensor, the $f(Q,T)$ gravity theory has been proposed \cite{Xu_2019_79,Bhagat_2023_42}. Recent studies have explored various aspects of $f(Q, T)$ and symmetric teleparallel gravity to understand the role of geometry and matter in cosmic evolution. Kaczmarek et al.\cite{Kaczmarek_2025} examined the scalar-tensor representation of $f(Q, T)$ gravity in a FLRW background, employing a suitable set of dynamical variables to cast the cosmological equations into an autonomous system. Najera et al. \cite{N_jera_2022} addressed the issue of degeneracy in the matter Lagrangian and analyzed its implications within this framework, highlighting how different formulations of matter coupling can influence the resulting field equations. Additionally, Sharif et al. \cite{Sharif_2024} investigated modified gravity in the context of ghost dark energy models, reconstructing the theory to study the influence of dark energy on late-time cosmic dynamics. This extension provides some crucial mechanism for the late time phenomena\cite{BHAGAT_2025_new,Narawade_2023_992}. \\

Another generalization of the $f(Q, T)$ gravity framework arises through its formulation in Weyl-type geometry \cite{Wheeler_2018}, where nonmetricity is represented by a vector field $\omega_{\mu}$, governed by the relation $\nabla_{\lambda} g_{\mu\nu} = -\omega_{\lambda} g_{\mu\nu}$. This geometrical approach modifies the gravitational action by including both kinetic and mass terms for $\omega_{\mu}$, thereby introducing a massive vector field coupled to the matter energy-momentum tensor \cite{Xu_2020_80}. To note, the divergence of energy-momentum tensor does not vanish, marking a departure from traditional GR and highlighting the influence of matter geometry coupling \cite{Gomes_2019}. Haghani et al. \cite{Haghani_2013_88} explored an extended Weyl$-$Cartan$-$Weitzenb$\ddot{o}$ck gravitational theory, incorporating the Weitzenb$\ddot{o}$ck condition via a Lagrange multiplier. Yang et al. \cite{Yang_2021_81} examined the geodesic deviation equation in this framework, providing insights into how nonmetricity affects particle trajectories. Bhagat et al. \cite{BHAGAT2025101913} investigated the cosmological implications of Weyl-type $f(Q, T)$ gravity by comparing theoretical predictions with observational data from various sources to constrain the model parameters. Recent developments in this direction, including works by Zhadyranova \cite{Zhadyranova_2024,Bhagat_ASPdyna2024}, further affirm the potential of this theory as a viable alternative to $\Lambda$CDM, offering a novel route \cite{Bhagat_2023_41}.\\

We investigate late-time issues with the Weyl-type $f(Q,T)$ gravity by considering two functional forms of $f(Q,T)$. To confront the models with the results of cosmological observations, we used cosmological datasets such as Cosmic Chronometers (CC), Pantheon$^+$ Type Ia supernovae, Union3.0, and DESI DR2. The paper is organized as follows. In Sec.--\ref{Sec:2}, the field equations of Weyl-type $f(Q,T)$ gravity has been presented. The data analysis description and framework have been given in Sec.--\ref{Sec:3}, where we apply the Markov Chain Monte Carlo (MCMC) technique to constrain the model parameters. In Sec.-- \ref{Sec:4}, we formulate both cosmological models and perform numerical integration to analyze the evolution of key cosmological quantities. Finally, Sec.-- \ref{Sec:5} summarizes our findings and highlights the implications of Weyl-type $f(Q,T)$ gravity as a viable alternative to standard cosmological models.\\

\section{Field Equations of Weyl type $f(Q,T)$ gravity}\label{Sec:2}
The action for Weyl-type $f(Q,T)$ gravity \cite{Xu_2020_80} given as,

\begin{eqnarray}\label{Eq:1}
S &=& \int \sqrt{-g} \big[\frac{1}{2} f(Q,T) -\frac{1}{4} W_{\mu\nu} W^{\mu\nu}-\frac{1}{2}m^{2} \omega_{\mu} \omega^{\mu}\nonumber\\&&
+\lambda (R +6\bigtriangledown_\alpha \omega^ \alpha-6 \omega_\alpha \omega^ \alpha) +\mathcal{L}_m \big]dx^4 ~,
\end{eqnarray}
where  $W_{\mu\nu} = \nabla_\mu \omega_\nu - \nabla_\nu \omega_\mu$ represents the antisymmetric field strength tensor of the Weyl vector field $\omega_\mu$.  $\mathcal{L}_m$ denotes the matter Lagrangian, and $m$ be the mass associated with the vector field. The second term in the action describes the standard kinetic contribution from the vector field, while the third term accounts for its mass. The parameter $\lambda$ is the Lagrange multiplier enforcing the Weyl condition in the geometric sector. Here we use the unit $8\pi G=c=1$.

The affine connection in Weyl geometry, known as the semi-metric connection, modifies the Levi-Civita connection and can be expressed as,

\begin{equation}\label{Eq:2}
    \tilde{\Gamma}^\lambda_{\mu\nu} = \Gamma^\lambda_{\mu\nu} + g_{\mu\nu} \omega^\lambda - \delta^\lambda_\mu \omega_\nu - \delta^\lambda_\nu \omega_\mu,
\end{equation}
where $\Gamma^\lambda_{\mu\nu}$ is the usual Christoffel symbol constructed from the metric $g_{\mu\nu}$. In Weyl geometry, the affine (semi-metric) connection modifies the Levi-Civita connection by allowing the metric to change under parallel transport according to $ \nabla_\lambda g_{\mu\nu} = -2 w_\lambda g_{\mu\nu} ,$ where $w_\lambda$ is the Weyl vector field responsible for simultaneous changes in direction and length. In contrast, the connection used in $f(Q)$ gravity, known as the symmetric teleparallel connection, is torsion-free and curvature-free but not metric compatible $\nabla_\lambda g_{\mu\nu} \neq 0$. Also, the nonmetricity tensor $Q_{\lambda\mu\nu} = \nabla_\lambda g_{\mu\nu}$ encoding how lengths change under parallel transport \cite{paliathanasisdyana,Basilakos_2025,ayuso2025}. Thus, while both geometries involve nonmetricity, Weyl geometry ties it to a single vector field $w_\lambda$, whereas $f(Q)$ gravity employs the full nonmetricity tensor to describe gravitational effects through the scalar $Q$.
 The nonmetricity tensor under this connection becomes

\begin{equation}\label{Eq:3}
Q_{\alpha\mu\nu} = \tilde{\nabla}_\alpha g_{\mu\nu} = \partial_\alpha g_{\mu\nu} - \tilde{\Gamma}^\rho_{\alpha\mu} g_{\rho\nu} - \tilde{\Gamma}^\rho_{\alpha\nu} g_{\rho\mu} = 2 \omega_\alpha g_{\mu\nu}.
\end{equation}

The scalar nonmetricity is defined as,
\begin{equation}\label{Eq:4}
Q = -g^{\mu\nu} \left( L^\alpha_{\beta\mu} L^\beta_{\nu\alpha} - L^\alpha_{\beta\alpha} L^\beta_{\mu\nu} \right),
\end{equation}
%
%where $L^\lambda_{\mu\nu}$ can be formed from the nonmetricity tensor as,
where 

\begin{equation}\label{Eq:5}
L^\lambda_{\mu\nu} = -\frac{1}{2} g^{\lambda\gamma} \left( Q_{\mu\gamma\nu} + Q_{\nu\gamma\mu} - Q_{\gamma\mu\nu} \right).
\end{equation}

Substituting $Q_{\alpha\mu\nu}$ into Eq.(\ref{Eq:5}), the scalar nonmetricity becomes,

\begin{equation}\label{Eq:6}
Q = -6 \omega^2,
\end{equation}
where $\omega^2 = \omega_\mu \omega^\mu$. Now varying the action with respect to the Weyl vector field $\omega_\mu$, the generalized Proca equation can be obtained as,

\begin{equation}\label{Eq:7}
    \nabla^\nu W_{\mu\nu} -  m_{\text{eff}}^2~\omega_\mu = 6 \nabla_\mu \lambda,
\end{equation}
where $m_{\text{eff}}=\left( m^2 + 6 f_Q + 12\lambda \right)^{1/2}$ is the effective mass of the Weyl vector field in a modified geometry scenario. Also, we denote the first order partial derivatives of $f(Q,T)$ with respect to $Q$ and $T$ respectively as $f_Q$ and $f_T$. One should note that, the presence of the term $f(Q,T)$ in the action, the mass of the Weyl field appears to be changed within the cosmic fluid. For brevity, we consider a unit mass for the Weyl field.

Applying the variational principle to the full action results, the field equations of Weyl-type $f(Q,T)$ gravity become,
\begin{multline}\label{Eq:8}
\frac{1}{2}\left[(1-f_T)T_{\mu\nu}+(S_{\mu\nu}-  f_T \zeta_{\mu\nu})\right]=-g_{\mu\nu}f\\-3 f_Q \omega_\mu \omega_\nu+\lambda(R_{\mu\nu}-6\omega_\mu \omega_\nu+3g_{\mu\nu}\triangledown_\rho \omega^\rho)\\+3g_{\mu\nu}\omega^\rho\triangledown_\rho\lambda-6\omega_{(\mu}Q_{\nu)}\lambda+g_{\mu\nu}\square \lambda- \triangledown_\mu\triangledown_\nu \lambda.
\end{multline}
 The energy-momentum tensor appearing in Eq. \eqref{Eq:8} is obtained from the matter Lagrangian $\mathcal{L}_m$ as

\begin{equation}\label{Eq:9}
    T_{\mu\nu} = -\frac{2}{\sqrt{-g}} \frac{\delta(\sqrt{-g}\mathcal{L}_m)}{\delta g^{\mu\nu}}.
\end{equation}
The additional tensor $\zeta_{\mu\nu}$ that arises due to the variation of $T_{\mu\nu}$ with respect to the metric can be expressed as
\begin{equation}\label{Eq:10}
\zeta_{\mu\nu} = g_{\mu\nu} \mathcal{L}_m - 2T_{\mu\nu} - 2g^{\alpha\beta} \frac{\delta^2 \mathcal{L}_m}{\delta g^{\mu\nu} \delta g^{\alpha\beta}}.
\end{equation}

The contributions to the energy-momentum tensor from the Weyl vector field may be expressed as
%In the modified field equations \eqref{Eq:8}, the tensor $S_{\mu\nu}$ represents the rescaled energy-momentum contributions from the Weyl vector field and is given as,

\begin{multline}\label{Eq:11}
S_{\mu\nu}=-\frac{1}{4}g_{\mu\nu}W_{\rho\sigma}W^{\rho\sigma}+W_{\mu\rho}W^\rho_\nu\\-\frac{1}{2}m^2 g_{\mu\nu}\omega_\rho \omega^\rho+m^2\omega_\mu \omega_\nu~.
\end{multline}

To construct the cosmological model of the Universe, we consider a spatially flat Friedmann-Lema\'itre-Robertson-Walker space time 

\begin{equation}\label{Eq:12}
ds^2 = -dt^2 + a^2(t)(dx^2 + dy^2 + dz^2),
\end{equation}
where $a(t)$ is the scale factor. The Hubble parameter relates the cosmic expansion as $H = \dot{a}/a$. Keeping the view of the spatial isotropy and homogeneity of the Universe, the Weyl vector should be time-like,

\begin{equation}\label{Eq:13}
\omega_\nu = [\varphi(t), 0, 0, 0].
\end{equation}
where $\varphi(t)$ is a function of time. Now we have, $\omega^2 =  -\varphi^2(t)$ and consequently, $Q = 6\varphi^2(t)$.  In comoving coordinates, the four-velocity vector  $u^\mu = (-1, 0, 0, 0)$ leads to $u^\mu \nabla_\mu = d/dt$. The matter Lagrangian is taken to be $\mathcal{L}_m = p$, giving rise to the energy-momentum tensor  $T^\mu_\nu = \text{diag}(-\rho, p, p, p)$ and  $\zeta^\mu_\nu = \text{diag}(2\rho + p, -p, -p, -p)$. 

The generalized Proca field equations governing the dynamics of the Weyl vector field then reduce to,

\begin{equation}\label{Eq:14}
    \dot{\varphi}-\varphi^2+3H\varphi-(\dot{H} + 2H^2) = 0, \quad \dot{\lambda} = -\frac{1}{6} m^2_{\text{eff}} \varphi, \quad \partial_i \lambda = 0.
\end{equation}
%
%
%where the effective mass of the vector field is given by $m_{\text{eff}} = \left(m^2 + 6 f_Q + 12\lambda\right)^{1/2}$.

The modified Friedmann equations in the Weyl-type $f(Q,T)$ gravity theory can be written as,
\begin{multline}\label{Eq:15}
\left(\frac{1+f_T}{2\lambda}\right)\rho+\left(\frac{f_T}{2\lambda}\right)p=\frac{f}{4\lambda}-\frac{1}{\lambda}\left(3f_Q+\frac{1}{4}m^2\right)\varphi^2 \\-3(\varphi^2-H^2)-3\frac{\dot\lambda}{\lambda}(\varphi-H),
\end{multline}
\begin{multline}\label{Eq:16}
-\frac{1}{2\lambda}p=\frac{1}{4\lambda}\left(f+m^2\varphi^2\right)+(3\varphi^2+3H^2+2\dot H)\\ +(3\varphi+2H)\frac{\dot\lambda}{\lambda}+\frac{\ddot\lambda}{\lambda} . 
\end{multline}

Using Eq. \eqref{Eq:14}, Eq. \eqref{Eq:15} and Eq. \eqref{Eq:16}, we obtain 
\begin{multline}\label{eq:17}
\frac{1}{2}\left(1+ f_T\right)\rho+\frac{1}{2}f_T p=\frac{f}{4}+\frac{m^2\varphi^2}{4}\\+3\lambda(H^2+\varphi^2)-\frac{1}{2}m^{2}_{eff}H\varphi,
\end{multline}
\begin{multline}\label{eq:18}
    \frac{1}{2}\left(1+f_T\right)(\rho+p)=\frac{m^{2}_{eff}}{6}\left(\dot{\varphi}+\varphi^2-H\varphi \right)\\+\dot{f_Q}\varphi-2\lambda\dot{H}.
\end{multline}
Substituting $\dot{\varphi}$ in Eq. \eqref{eq:18}, we obtain 
\begin{multline}\label{eq:19}
    \frac{1}{2}\left(1+f_T\right)(\rho+p)=-2\lambda \Bigl(1-\frac{m^{2}_{eff}}{12\lambda}\Bigl)\dot{H}\\+\frac{m^{2}_{eff}}{3}\left(H^2+\varphi^2-2H\varphi \right)+\dot{f_Q}\varphi.
\end{multline}
The equation for the energy balance is,
\begin{equation}\label{eq:20}
    \dot{\rho}+3H(\rho+p)=\frac{1}{1+f_{T}}((\rho+p)\dot{f_T}-f_T(\dot{\rho}-\dot{p})).
\end{equation}

When $f_T=0$ the modified balance equation vanishes and we recover the standard conservation law $\dot{\rho}+3H(\rho+p)=0$; therefore non-conservation of matter arises only when the model includes a nonzero coupling $f_T$.

Reformulating the generalized Friedmann equations for an effective fluid description as
\begin{eqnarray}
3H^2&=&\frac{1}{2\lambda}(\rho+\rho_{d}),\label{eq:17}\\
3H^2+2\dot{H}&=&-\frac{1}{2\lambda}(p+p_{d}), \label{eq:18}
\end{eqnarray}
we may get the effective additional components of the energy density and pressure as
%where the effective energy density and pressure are respectively,
\begin{multline}\label{eq:19}
\rho_{d}=m^{2}_{\text{eff}}H\varphi+f_T(\rho+p)-\frac{f}{2} -\frac{{\varphi^2}}{2}-6\lambda\varphi^2,
\end{multline}
\begin{multline}\label{eq:20}
p_{d}=\frac{m^{2}_{\text{eff}}}{3}(\dot{\varphi}+\varphi^2-4H\varphi)+\frac{f}{2}+2\dot{f_Q}\varphi  \\ +\frac{\varphi^2}{2}+6\lambda\varphi^2.
\end{multline}

One should note that, the total energy density includes the usual energy density and the contribution coming from the geometry modification and is expressed as $\rho_{tot}=\frac{1}{2\lambda}(\rho+\rho_{d})$. Similarly total pressure becomes $p_{tot}=\frac{1}{2\lambda}(p+p_{d})$. The equation of state parameter for the effective fluid description becomes
\begin{equation}
    w = \frac{p+p_d}{\rho+\rho_d}.
\end{equation}

In order to assess the cosmological implication of the Weyl type $f(Q,T)$ gravity theory, it is required to consider specific functional forms of $f(Q,T)$ and constrained the models using different observational data sets. In the following sections, we will employ numerical techniques to solve the system of equations for the models and discuss the cosmological data sets  to constrain the model parameters.

\section{Cosmological Datasets}\label{Sec:3} 

In this section, we briefly discuss the cosmological datasets to be used to frame the cosmological model. The Pantheon+ sample of SNIa data \cite{Brout_2022_938}, Cosmic Chronometers (CC) data \cite{Moresco_2022_25,Jimenez_2002_573}, Union3.0 \cite{rubin20logy2000} and DESI DR2 data sets\cite{Alam_2004_2004_008} are used for the purpose. We employ the \texttt{emcee} package for an MCMC simulation \cite{Foreman-Mackey_2013_125}, which will enable to constrain both the model and cosmological parameters, thereby examining the parameter space's posterior.

The {\bf CC dataset} provides an essential method to directly constrain the Hubble rate $H(z)$ across different redshifts. We used thirty-two data points compiled from Refs. \cite{Stern_2010_2010_008,Borghi_2022_928}. 

In MCMC analysis, we used \( \chi^2_{\text{CC}} \) to assess the agreement between the theoretical Hubble parameter values \( H(z_i,\Theta) \), given the model parameters \( \Theta \), and the observed Hubble data values \( H_{\text{obs}}(z_i) \), with an observational error \( \sigma_H(z_i) \). The \( \chi^2_{\text{CC}} \) was calculated using the following formula,

\begin{equation}
\chi^2_{\text{CC}} = \sum_{i=1}^{32} \frac{(H(z_i,\Theta) - H_{\text{obs}}(z_i))^2}{\sigma^2_H(z_i)}.
\end{equation}

The {\bf  Pantheon$^+$}  sample \cite{Brout_2022_938} is one of the most extensive collections of SNIa data, comprising $1701$ measurements that cover a redshift range of 0.0 $<$ z $<$ 2.26. This determines the expansion rate of the Universe, $H(z)$, by measuring the observed apparent magnitude ($m$) and absolute magnitude ($M$).
\begin{equation}
\mu(z_i, \Theta) = m - M = 5 \log_{10} [D_L(z_i, \Theta)] + 25,
\end{equation}
where $\Theta$ represents the set of cosmological parameters that describe the Universe, $z_i$ is the redshift of the SNIa measurement. The luminosity distance $D_L(z_i, \Theta)$ can be expressed as,

\begin{equation}
D_L(z_i, \Theta) = (1 + z_i) \int_0^{z_i} \frac{dz'}{H(z', \Theta)}.
\end{equation}
%$c$ be the speed of light. 
To constrain the cosmological parameters, the minimum $\chi^2$ can be calculated as,

\begin{equation}
\chi^2_{\text{SN}} = (\Delta \mu(z_i, \Theta))^T C^{-1}_{\text{SN}} (\Delta \mu(z_i, \Theta)),
\end{equation}
where $\Delta \mu(z_i, \Theta) = \mu(z_i, \Theta) - \mu(z_i)_{\text{obs}}$. $C_{\text{SN}}$ is the corresponding covariance matrix that accounts for the statistical and systematic uncertainties.\\

The {\bf Union3.0} Supernova compilation consists of 2,087 Type Ia supernovae collected from 24 different surveys \cite{rubin20logy2000}, spanning a redshift range of 0.01 to 2.26. These supernovae have been calibrated to a consistent distance scale using the SALT3 light-curve fitter. The dataset has been processed and binned with the UNITY1.5 Bayesian framework, and a total of 22 binned distance measurements from Union3.0. For the Union3.0 supernova dataset, the chi-squared is computed as
\begin{equation}
    \chi^2 = \sum_{i=1}^{22} \frac{\left[\mu_i^{\rm obs} - \mu_i^{\rm th}(z_i, \boldsymbol{\theta})\right]^2}{\sigma_i^2},
\end{equation}

where $\mu_i^{\rm obs}$ and $\sigma_i$ are the observed distance moduli and uncertainties, and $\mu_i^{\rm th}$ is the theoretical value for model parameters $\boldsymbol{\theta}$. The minimum $\chi^2_{\rm min}$ is obtained by varying the parameters to best fit the data.

The {\bf DESI DR2} data release \cite{rubin20logy2000,desicollaboration2025desidr2resultsii}. The analysis provides either two correlated distance ratios, $D_M/r_d$ and $D_H/r_d$, or a single ratio, $D_V/r_d$. Here, $r_d$ denotes the sound horizon at the drag epoch. The distance measures are expressed as
\begin{align}
D_H(z) &= \frac{c}{H(z)}, \\
D_M(z) &= \frac{c}{H_0} \int_0^z \frac{dz_*}{E(z_*)}, \\
D_V(z) &= \left[ z\, D_H(z)\, D_M^2(z) \right]^{1/3}.
\end{align}

These measurements are derived from multiple redshift ranges using different tracers, including bright galaxies (BGS), luminous red galaxies (LRG), emission line galaxies (ELG), quasars (QSO), and the Lyman-$\alpha$ forest (Ly$\alpha$) from high-redshift quasars.
 For the DESI BAO data, the chi-squared is computed as
 \begin{equation}
     \chi^2 = \sum_{i,j} \left(D_i^{\rm obs} - D_i^{\rm th}\right) (C^{-1})_{ij} \left(D_j^{\rm obs} - D_j^{\rm th}\right),
 \end{equation}
where $D_i^{\rm obs}$ and $D_i^{\rm th}$ are the observed and theoretical BAO distance measures, and $C^{-1}$ is the inverse covariance matrix. The minimum chi-squared, $\chi^2_{\rm min}$, is obtained by varying the model parameters to best fit the data.

To assess the model while accounting for its complexity, we compute the Akaike Information Criterion (AIC) \cite{Akaike_1974} and Bayesian Information Criterion (BIC) \cite{Schwarz_1978}, which balance goodness of fit against the number of free parameters. The AIC is given by
\begin{equation}
    \text{AIC} = \chi^2_{\rm min} + 2k.
\end{equation}
The statistical distinction between two models can be assessed using the difference in their AIC values, defined as
\begin{equation}
\Delta\text{AIC} = \text{AIC} - \text{AIC}_{\Lambda},  
\end{equation}

which indicates how strongly one model is favored over the other. 

and the BIC is
\begin{equation}
    \text{BIC} = \chi^2_{\rm min} + k \ln(N).
\end{equation}
The difference in the BIC values between two models is given by
\begin{equation}
   \Delta\text{BIC} = \text{BIC} - \text{BIC}_{\Lambda}, 
\end{equation}

which we have analyzed in our study to compare both models statistically.

Here $\chi^2_{\rm min}$ is the minimum chi-squared, $k$ is the number of free parameters, and $N$ is the number of data points. To statistically compare our model with the $\Lambda$CDM model, we first present the results obtained using the CC+Union3.0+DESI DR2 and CC+Pantheon$^+$+DESI DR2 datasets in \hyperref[Appendix-I]{Appendix-I}).

To establish a connection between theoretical predictions and observational data in cosmology, it is standard to replace the time variable $t$ with the redshift parameter $z$, which is more directly accessible through astrophysical observations. The scale factor $a$ and redshift $z$ are related through $ z = \frac{1}{a}-1$, where the present day scale factor is considered as $a(0) = 1$. In the analysis, we use 
\begin{align}
\frac{d}{dt} = \frac{dz}{dt} \frac{d}{dz} = -(1+z)H(z)\frac{d}{dz},
\end{align}
which allows us to express the cosmological quantities as a function of redshift. 

\section{The Cosmological Models} \label{Sec:4}
In this section, we present two cosmological models based on different choices of the functional form of $f(Q,T)$.\\

\vspace{0.2in}
{\bf Model I: Logarithmic model}

As a first choice, we consider the modified gravitational Lagrangian as a logarithmic function \cite{BHAGAT2026100483,bhagat2025eff} of the nonmetricity scalar $Q$, along with a linear contribution from the trace of the energy-momentum tensor $T$ as
\begin{equation}
    f(Q, T) = Q \log\left(\alpha \frac{Q_0}{Q} \right) + \beta T,
\end{equation}
where $\alpha$ and $\beta$ are model parameters and $Q_0=6\varphi_0^2$, with $\varphi_0$ being the initial value of $\varphi$.
The logarithmic form of $f(Q)$ has been explored in recent literature as a viable modification of symmetric teleparallel gravity. Unlike the standard linear dependence on $Q$ in the coincident GR limit, the logarithmic contribution introduces a nontrivial modification to the effective gravitational coupling at different curvature (or, in this framework, nonmetricity) scales. Such forms are motivated by their ability to naturally generate late-time cosmic acceleration without explicitly introducing a cosmological constant \cite{Narawade_2023_992}, thereby offering a possible resolution to the cosmological constant problem.
Logarithmic corrections often appear in effective field theories. This choice provides rich phenomenology, since the term $Q \log(Q)$ tends to suppress the nonmetricity contribution at high $Q$ (early universe), while enhancing deviations from GR at low $Q$ (late-time universe). This behavior makes the model suitable for describing both the matter-dominated era and the subsequent accelerated expansion in a unified framework.
The additional linear coupling with the trace of the energy-momentum tensor, $\beta T$, introduces matter geometry interaction. This interaction can lead to non-conservation of the standard energy momentum tensor, thereby opening new possibilities for explaining dynamical dark energy effects.

{\bf Model II: Strong coupling model} 

We consider another form for $f(Q,T)$ as
\begin{equation}
f(Q, T) = \frac{\gamma}{3H_0^2}QT,
\end{equation}
where $\gamma$ is a constant model parameter. This form of $f(Q,T)$ adopts an involved interaction between the nonmetricity scalar $Q$ and the trace of the energy-momentum tensor $T$ \cite{Xu_2019_79}.

The interaction term modifies the evolution equations by coupling the geometry directly to the matter sector. This interaction influences the evolution of the Hubble parameter $H(z)$, the scalar field $\varphi(z)$, the auxiliary function $\lambda(z)$, and the density $\rho$. The partial derivative $f_Q = \tfrac{\gamma}{3H_0^2} T$ in this model explicitly incorporates the energy density into the field equations, leading to different dynamical behavior.

Such a cross-coupling between $Q$ and $T$ represents a non-minimal interaction model in the $f(Q,T)$ framework. Unlike additive models $f(Q,T) = f_1(Q) + f_2(T)$, the multiplicative $QT$ form generates an effective feedback mechanism where the dynamics of the geometry and matter sectors are inseparably linked. As a result, the usual conservation law of the energy momentum tensor does not hold, i.e., $\nabla_\mu T^{\mu\nu} \neq 0$, implying an exchange of energy between matter and the effective dark energy sector.

From the generalized Proca equation and the modified Friedmann equations, we derive a system of first-order differential equations ( given in the \hyperref[Appendix-II]{Appendix-II}) governing the evolution of the Hubble parameter $H(z)$, the scalar field $\varphi(z)$, the auxiliary function $\lambda(z)$ and the density $\rho$. Due to the logarithmic nature or the strong coupling of this system, analytical solutions are not feasible. Therefore, we employ numerical integration techniques to simultaneously solve the system and analyze the evolution of key cosmological quantities. In our analysis, we assume that the Universe is primarily composed of pressureless dust matter with $p = 0$ and the mass of the Weyl field $m$, as the unit of mass. This serves as a suitable approximation for the late-time evolution of the cosmos, where non-relativistic matter dominates the overall dynamics. To constrain the free model parameters $\alpha$, $\beta$, $\gamma$ along with initial conditions $\varphi_0$, $\lambda_0$, and present expansion rate $H_0$, we perform a Bayesian parameter estimation using MCMC methods. This analysis incorporates observational data from three independent cosmological probes: the CC, the Pantheon$^+$ compilation of Type Ia supernovae and BAO).\\
%

%  \begin{figure}[H]
%     \centering
%     \includegraphics[width=7.5cm,height=7.5cm]{Contour_Log.pdf}
%     \includegraphics[width=7.5cm,height=7.5cm]{Contour_QT.pdf}

%     \caption{ The contour plots for the Logarithmic (Upper Panel) and the Strong coupling (Lower Panel) model with 32 points of CC sample, 1701 light curves from Pantheon$^+$ dataset and 6-BAO data upto 2$-\sigma$ errors for the parameters.
% } 
%     \label{Fig1}
% \end{figure} 
\begin{widetext}
   
 \begin{figure}[H]
    \centering
    \includegraphics[width=7.5cm,height=7.5cm]{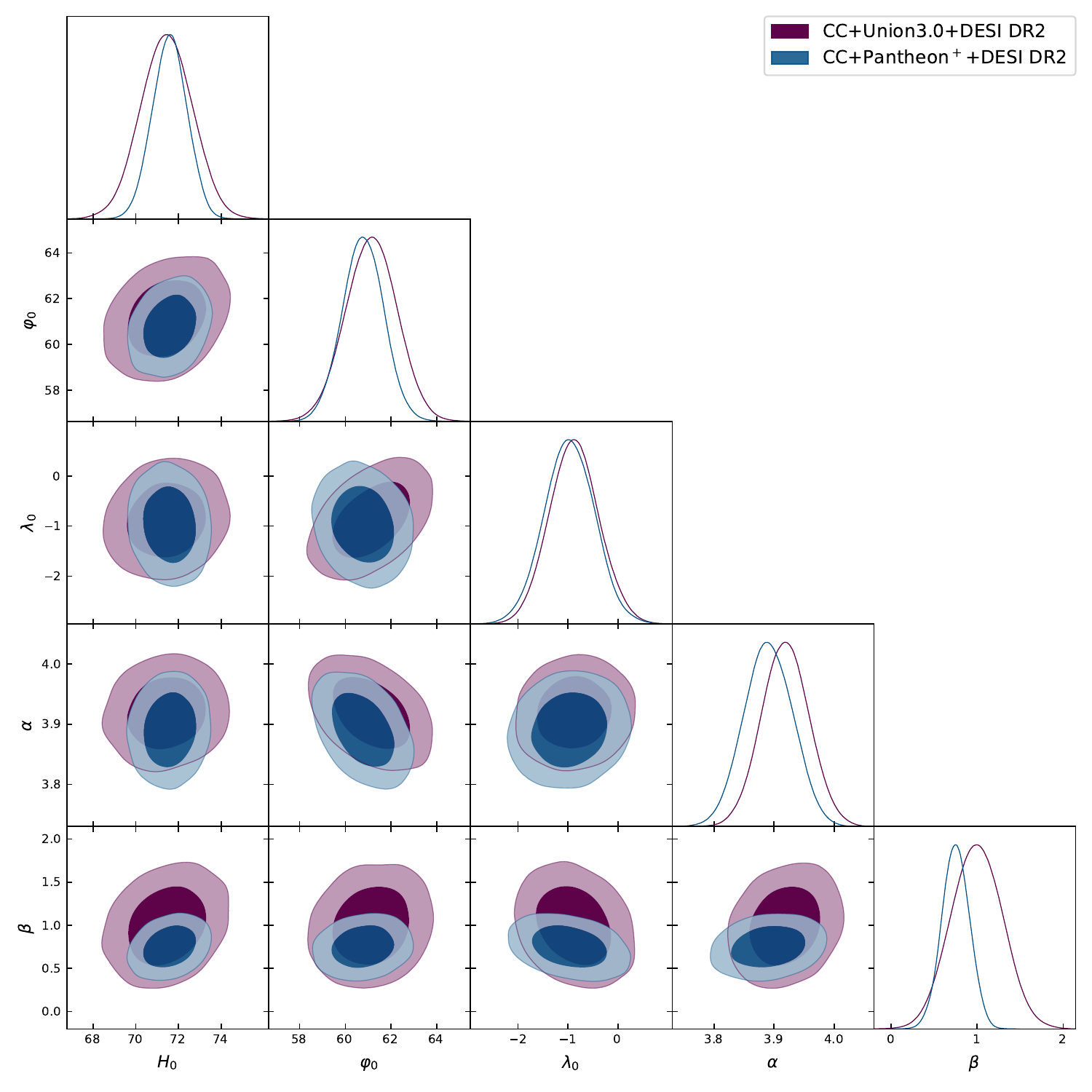}
    \includegraphics[width=7.5cm,height=7.5cm]{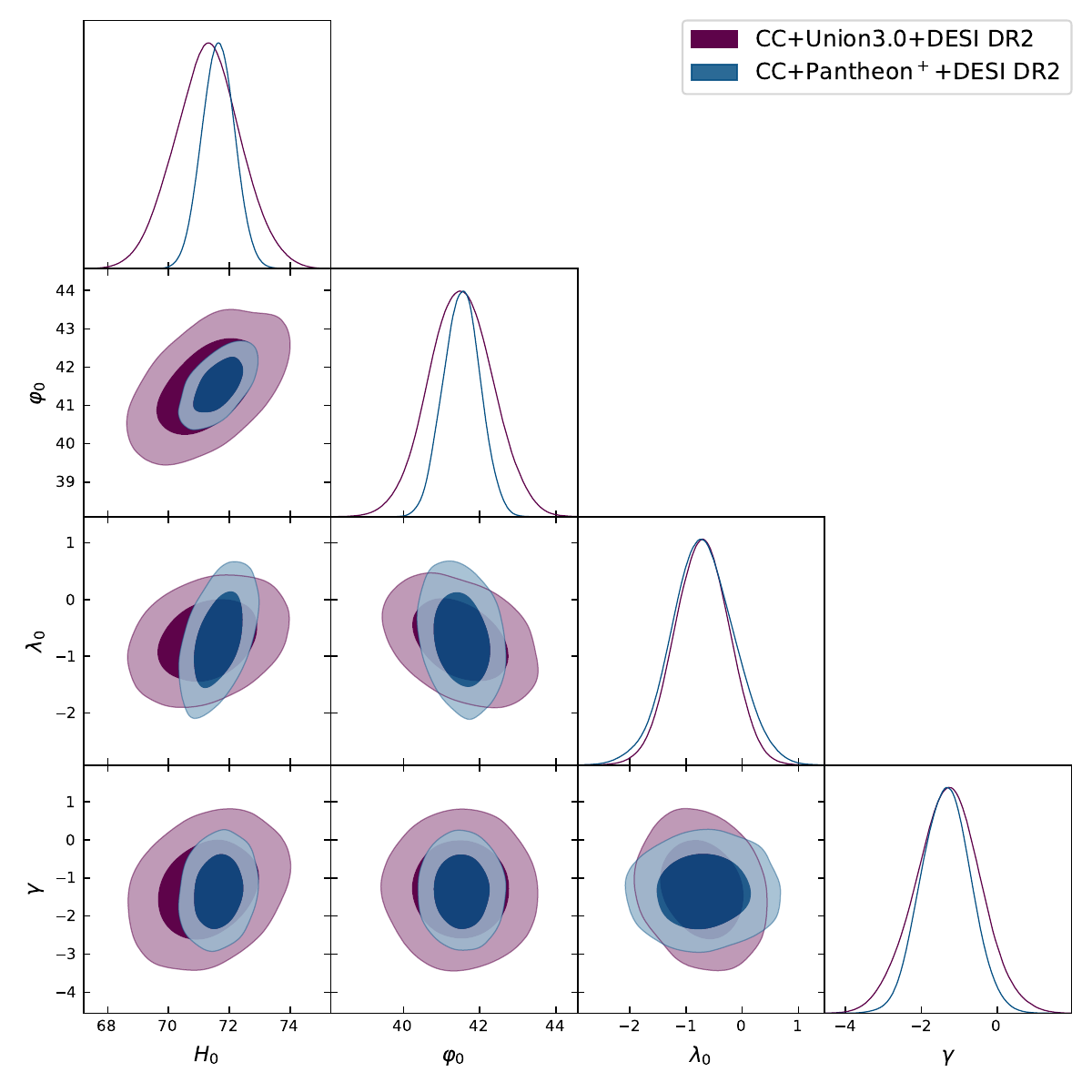}

    \caption{ The contour plots for the Logarithmic (Left Panel) and the Strong coupling (Right Panel) model CC dataset, Union3.0, Pantheon$^+$ dataset and DESI DR2 dataset data upto 2$-\sigma$ errors for the parameters.
} 
    \label{Fig1}
\end{figure} 

\begin{figure}[H]
\centering
\includegraphics[width=8cm,height=5.5cm]{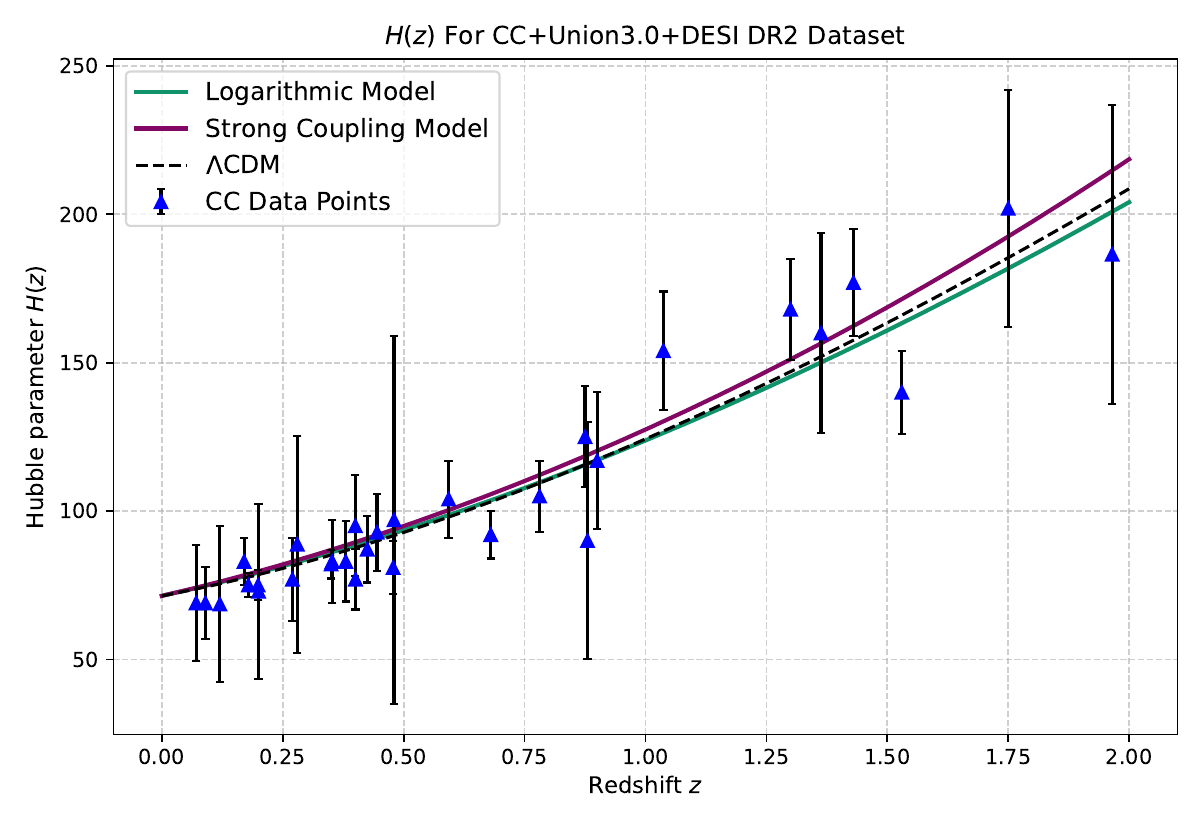}
\includegraphics[width=8cm,height=5.5cm]{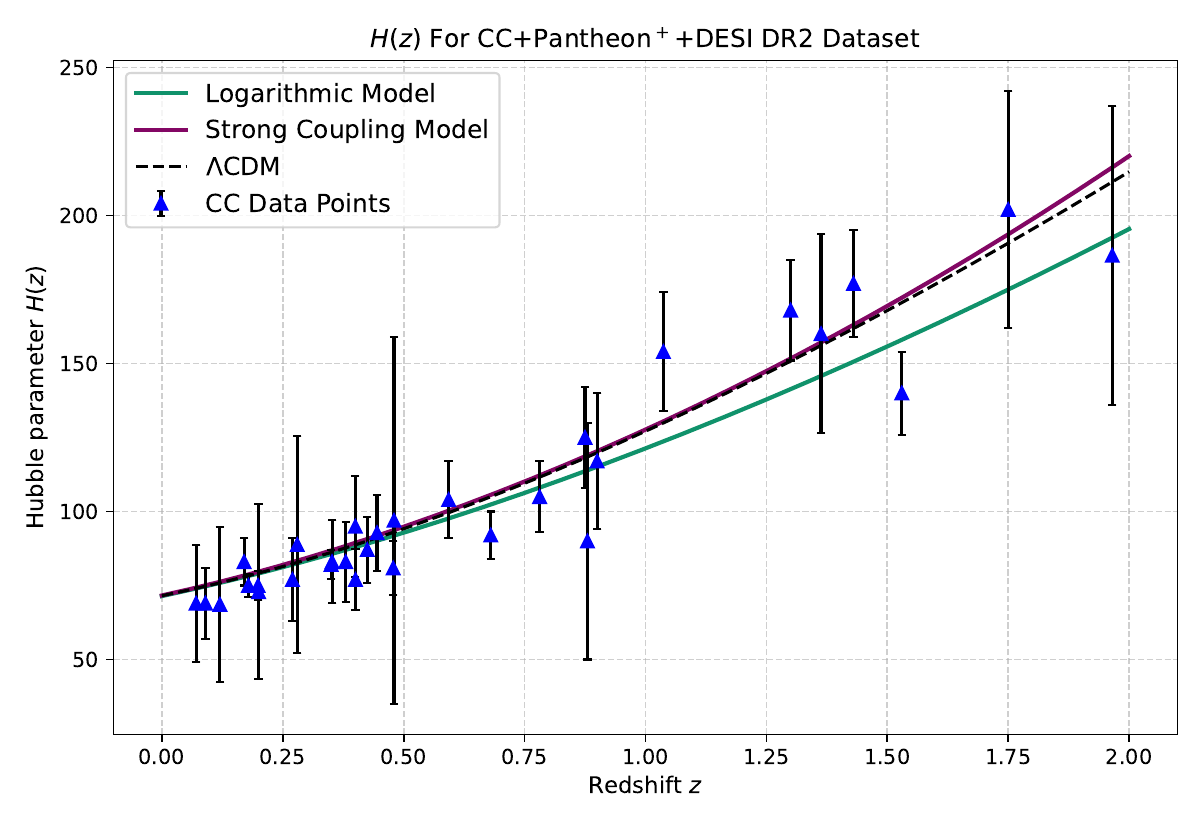}
\caption{{Evolution of Hubble parameter for both the model.}}
\label{Fig2}
\end{figure}
\begin{table*}[htp]
\renewcommand\arraystretch{1.5}
\centering % used for centering table

\begin{tabular}{|c|c|c|c|c|c|c|c|c|c|} % centered columns (3 columns)
\hline %inserts double horizontal lines
~~~ \parbox[c][1.3cm]{2cm}{Logarithmic Model Parameters} ~~~& ~~~$H_0$  ~~~&~~~~$\varphi_{0}$~~~~&~~~~$\lambda_{0}$~~~~&~~~~$\alpha$~~~~ & ~~~$\beta$~~~&    \parbox[c][1cm]{1.8cm}{$\chi^2_{min}$ - $\chi^2_{\Lambda~min}$} &$\Delta\text{AIC}$&$\Delta\text{BIC}$
     \\ [0.5ex] % inserts table
%heading
% inserts single horizontal line
\hline\hline
CC+Union3.0+DESI DR2  & $71.43\pm1.22$ &  $61.149^{+1.125}_{-1.173}$ & $-0.872\pm0.495$&$3.918^{+0.039}_{-0.034}$&$0.994\pm0.304$& $4.36$&  $10.36$&$16.929$ \\
\hline
CC+Pantheon$^+$+DESI DR2  & $71.58^{+0.81}_{-0.79}$ &  $60.794^{+0.915}_{-0.927}$ & $-0.983^{+0.624}_{-0.504}$&$3.891\pm0.042$&$0.751\pm0.165$& $5.89$&  $11.89$&$28.283$ \\
\hline
\end{tabular}
\caption{{Constrained parameter values obtained for joint datasets for the Logarithmic Weyl type $f(Q, T)$ gravity model.}} 
\label{tableA1}
\end{table*}
\end{widetext}

The marginalized posterior distributions and parameter correlations are illustrated through dimensional contour plots in Fig \ref{Fig1}, representing the 68$\%$ and 95$\%$ confidence levels. These plots reveal the allowed regions in the multidimensional parameter space and help visualize the degeneracies among the parameters. The corresponding best-fit values and uncertainties obtained from the MCMC chains are summarized in Table \ref{tableA1} and \ref{tableA2}. These serve as the basis for all subsequent theoretical reconstructions and observational comparisons.

To analyze the expansion history predicted by each model, we reconstruct the redshift evolution of the Hubble parameter $H(z)$. In Fig. \ref{Fig2} (Right Panel), for CC+Union3.0+DESI DR2, we compare the $H(z)$ curves derived from both models with the $\Lambda$CDM benchmark, using the best-fit parameters from our MCMC analysis. We also include the 32 data points from the CC dataset for direct comparison. Both models show excellent agreement with $\Lambda$CDM at low redshifts, accurately fitting the observational data. We have also examined the difference in the minimum chi-square values $\Delta$$\chi^2_{\text{min}}$ for both models, along with their corresponding $\Delta$AIC and $\Delta$BIC values, to assess their comparative statistical performance in Table \ref{tableA1} and \ref{tableA2}. However, deviations appear at higher redshifts ($z \gtrsim 2$), where the strong coupling model predicts a more rapid expansion compared to $\Lambda$CDM.

\begin{table*}[htp]
\renewcommand\arraystretch{1.5}
\centering % used for centering table
{
\begin{tabular}{|c|c|c|c|c|c|c|c|c|c|} % centered columns (3 columns)
\hline %inserts double horizontal lines
~~~ Strong Coupling Model Parameters ~~~& ~~~$H_0$  ~~~&~~~~$\varphi_{0}$~~~~&~~~~$\lambda_{0}$~~~~&~~~~$\gamma$~~~~ &    \parbox[c][1cm]{1.8cm}{$\chi^2_{min}$ - $\chi^2_{\Lambda~min}$} &$\Delta\text{AIC}$&$\Delta\text{BIC}$
     \\ [0.5ex] % inserts table
%heading
% inserts single horizontal line
\hline\hline
CC+Union3.0+DESI DR2  & $71.39\pm1.17$ &  $41.492\pm0.843$ & $-0.713^{+0.521}_{-0.457}$&$-1.297^{+0.845}_{-0.864}$& $5.23$&  $9.23$&$13.609$ \\
\hline
CC+Pantheon$^+$+DESI DR2  & $71.64^{+0.54}_{-0.53}$ &  $41.526\pm0.485$ & $-0.685^{+0.561}_{-0.553}$&$-1.362^{+0.658}_{-0.649}$& $4.41$&  $8.41$&$19.339$ \\
\hline
\end{tabular}}
\caption{{ Constrained parameter values obtained for joint datasets for the Strong coupling model.}} 
\label{tableA2}
\end{table*}

\begin{widetext}
\begin{figure}[H]
\centering
\includegraphics[width=19cm]{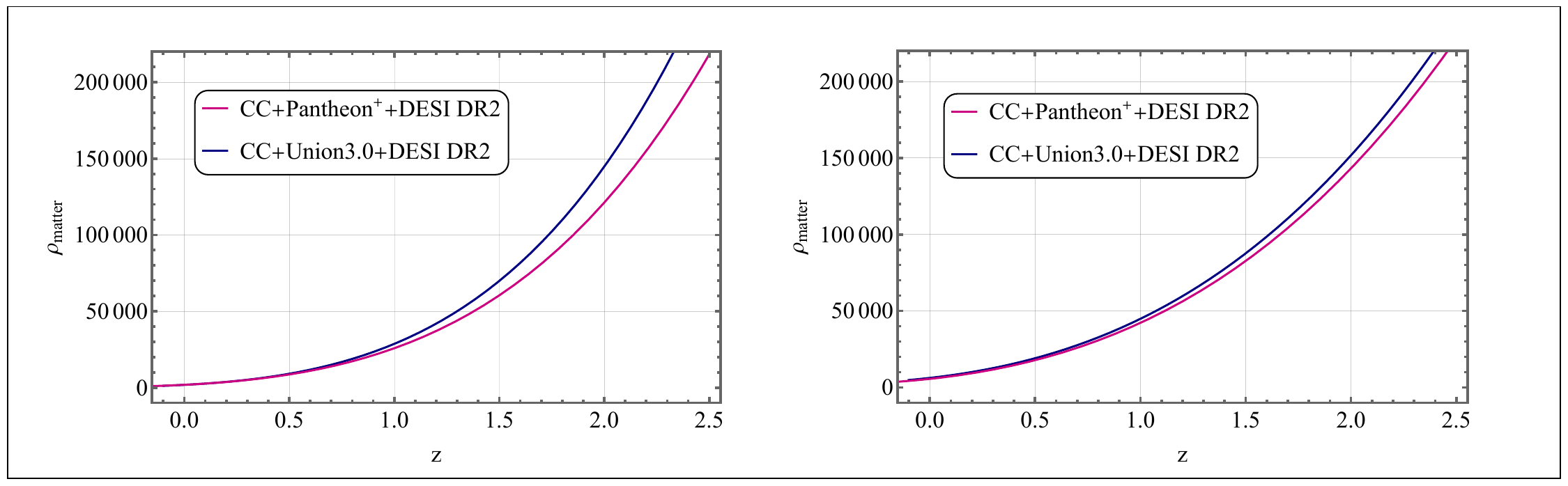}
\caption{{Behavior of matter energy density. Logarithmic: Left Panel, Strong Coupling: Right Panel.}}
\label{Fig3}
\end{figure}
\end{widetext}

To further investigate the cosmological implications of both models, we examine and compare the evolution of key cosmological parameters. Fig. \ref{Fig3} displays the behavior of the matter energy density $\rho_{matter}$ for the Logarithmic and the Strong coupling  models. In both cases, the total energy density remains positive throughout cosmic evolution and gradually approaches zero at late times, consistent with expectations from standard cosmology.

\begin{widetext}
    \begin{figure}[H]
\centering
\includegraphics[width=19cm]{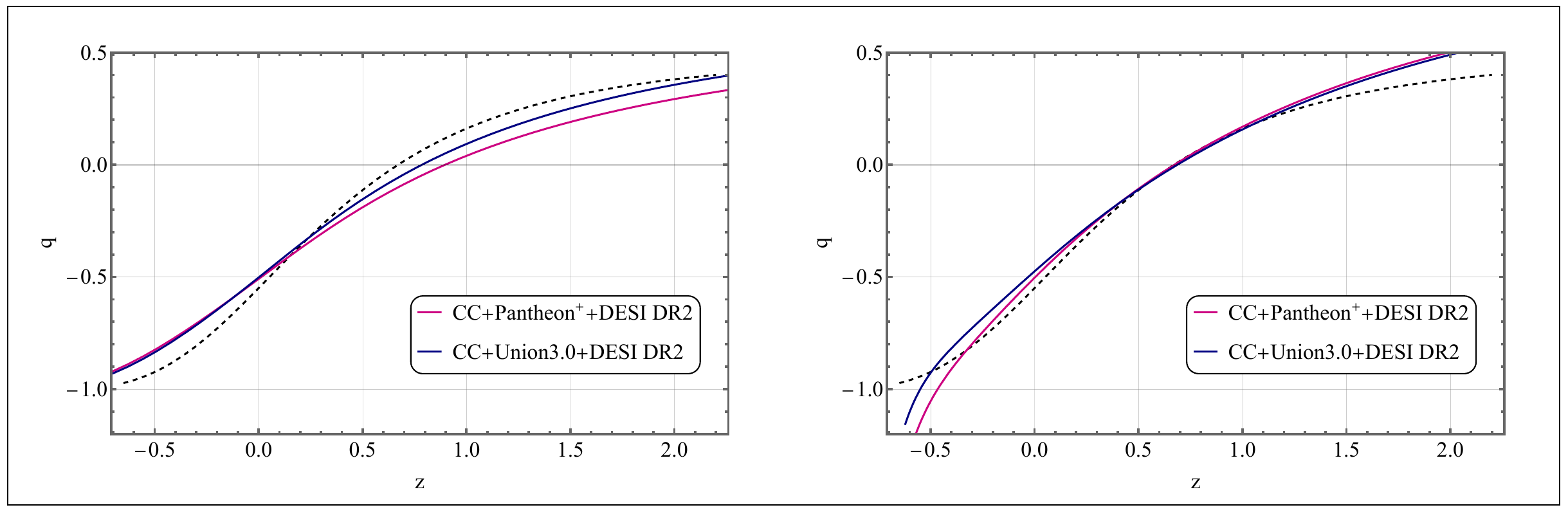}
\caption{{Behavior of Deceleration Parameter with redshift. Logarithmic: Left Panel, Strong Coupling: Right Panel.}}
\label{Fig4}
\end{figure}
\end{widetext}

The deceleration parameter $q(z)$, presented in Fig. \ref{Fig4}, further highlights the dynamical differences between the models. The Logarithmic model closely tracks the $\Lambda$CDM behavior, asymptotically approaching $q = -1$ at late times, which signifies an era of sustained accelerated expansion. In contrast, the Strong Coupling model shows a similar deceleration trend in the low-redshift range ($0 < z < 1$), but it deviates at higher redshifts owing to its intrinsic non-linear nature. Moreover, it exhibits a phantom-like behavior for $z < 0$, indicating a possible super-accelerated future expansion. This divergence suggests a modified expansion history at early epochs. Nevertheless, both models produce nearly identical present-day values of the deceleration parameter, as shown in Table \ref{table:3}. The transition redshift from deceleration to acceleration is found to be slightly higher ($z_{tr}=0.785$ and $z_{tr}=0.897$) for the Logarithmic model, while the Strong Coupling model closely aligns with the $\Lambda$CDM behavior for both combined datasets: CC+Union3.0+DESI DR2 and CC+Pantheon$^+$+DESI DR2, respectively. These results are in good agreement with observational bounds reported in recent literature \cite{Yang_2020_2020_059,Capozziello_2014_90_044016}.

\
\begin{widetext}
 \begin{table}[htb]
\renewcommand\arraystretch{1.5}
\centering % used for centering table
{
\begin{tabular}{|c|c|c|c|c|}
\hline
Cosmological Parameter & \parbox[c][1cm]{3.5cm}{CC+Union3.0+DESI DR2 for Model-I} & \parbox[c][1cm]{3.5cm}{ CC+Pantheon$^+$+DESI DR2 for Model-I}& \parbox[c][1cm]{3.5cm}{CC+Union3.0+DESI DR2 for Model-II} &  \parbox[c][1cm]{3.5cm}{CC+Pantheon$^+$+DESI DR2 for Model-II}\\
\hline
$q_0$  & $-0.502$ & $-0.509$ & $-0.475$ & $-0.503$ \\
\hline
$z_{\mathrm{tr}}$ & $0.785$ & $0.897$ & $0.687$ & $0.673$ \\
\hline
$w_0$  & $-0.668$ & $-0.672$ & $-0.650$ & $-0.668$ \\
\hline
$s_0$  & $0.084$ & $0.100$ & $0.070$ & $0.025$ \\
\hline
$r_0$  & $0.744$ & $0.692$ & $0.793$ & $0.924$ \\
\hline
Age of the Universe [Gyr] & $14.013$ & $13.893$ & $13.818$ & $13.739$ \\
\hline
\end{tabular}}
\caption{{Present values of the cosmological parameters.}}
\label{table:3}
\end{table}
\end{widetext}

\begin{widetext}
    \begin{figure}[H]
\centering
\includegraphics[width=19cm]{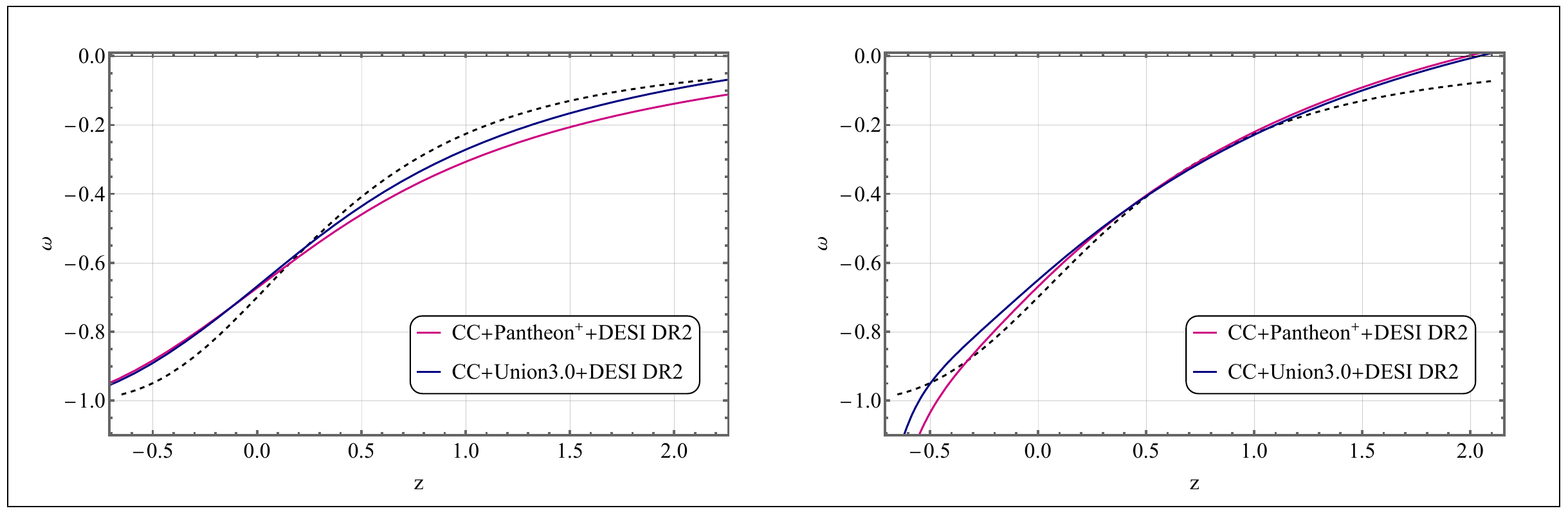}
\caption{{Behavior of Equation of State Parameter with redshift. Logarithmic: Left Panel, Strong Coupling: Right Panel.}}
\label{Fig5}
\end{figure}
\end{widetext}

In addition, we explore the evolution of the equation of state (EoS) parameter $w(z)$, which characterizes the nature of the dominant energy component driving cosmic expansion. As shown in Fig. \ref{Fig5}, both models yield a present-day value of $w_0 \sim -0.6$, indicating a dark energy component that drives accelerated expansion but is slightly less negative than the cosmological constant ($w = -1$). At lower redshifts, $w(z)$ for both models evolves smoothly. While $w(z)$ for the logarithmic model tends to approach $-1$ at late times, $w(z)$ for the Strong coupling model goes below the phantom divide. These results are consistent with the expected behavior for models mimicking $\Lambda$CDM at late times. This transition reflects the model capability to replicate the observed accelerated expansion while allowing for more flexibility in early-Universe dynamics.

An additional diagnostic used to assess the nature of dark energy is the $Om(z)$ parameter \cite{Sahni_2008_78_103502}, which provides a model-independent method to distinguish between various cosmological scenarios. Defined as a function of the Hubble parameter $H(z)$ and redshift $z$, the $Om(z)$ diagnostic is given as,

\begin{equation}
    Om(z) = \frac{H^2(z)/H_0^2 - 1}{(1+z)^3 - 1}.
\end{equation}

In the standard $\Lambda$CDM model, where dark energy is characterized by a cosmological constant ($w = -1$), the $Om(z)$ parameter is expected to remain constant across redshifts. Any deviation from this constant behavior suggests a departure from the cosmological constant scenario and hints at a dynamical dark energy component. The slope of $Om(z)$ is particularly useful in this context: a positive slope implies a phantom dark energy regime ($w < -1$), while a negative slope corresponds to a quintessence-like behavior ($w > -1$), where dark energy evolves over time.

\begin{widetext}
    \begin{figure}[H]
\centering
\includegraphics[width=19cm]{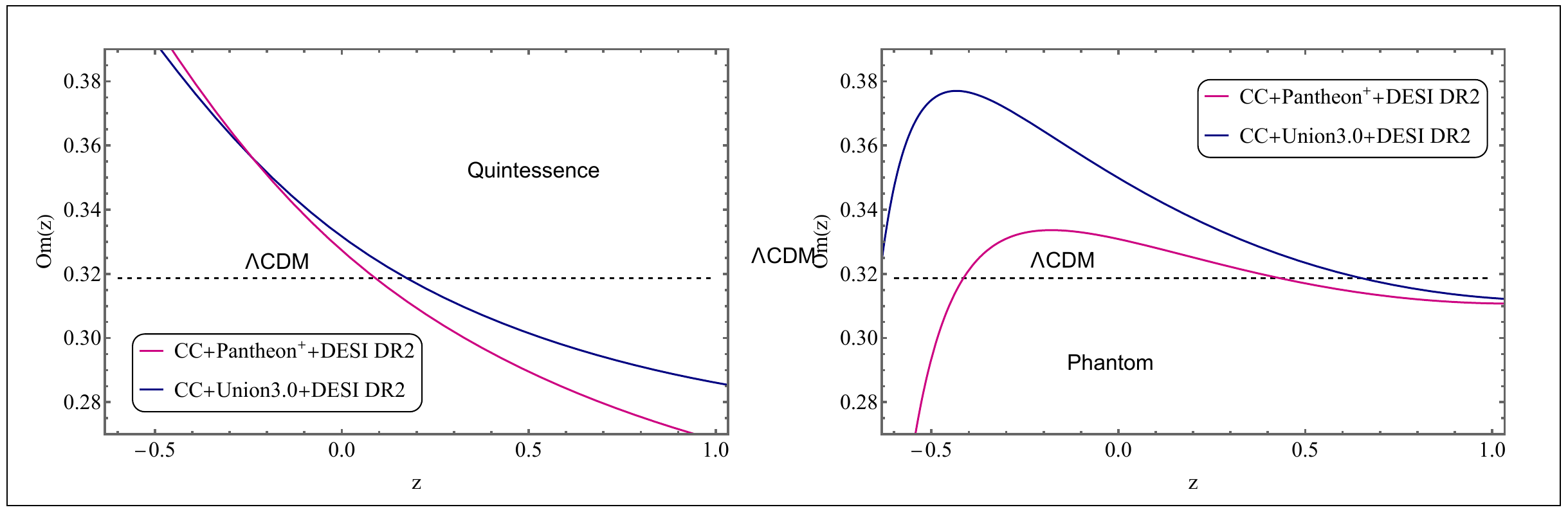}
\caption{{Behavior of $Om(z)$ Parameter with redshift. Logarithmic: Left Panel, Strong Coupling: Right Panel.}}
\label{Fig6}
\end{figure}
\end{widetext}

In our analysis, we evaluate the $Om(z)$ parameter for both the Logarithmic and Strong Coupling models using the best-fit parameters obtained from the MCMC analysis. As shown in Fig.~\ref{Fig6}, the Logarithmic model exhibits a negative slope in $Om(z)$, indicating quintessence-like behavior for both datasets. In contrast, the Strong Coupling model shows a negative slope at positive redshifts (quintessence behavior) and a positive slope at negative redshifts, corresponding to phantom behavior. This phantom feature is further supported by the evolution of the deceleration parameter, where the dotted black line represents the $\Lambda$CDM reference model. This behavior supports our earlier findings derived from the equation of state parameter $w(z)$. The negative slope of $Om(z)$ serves as independent evidence of quintessence-like dynamics and reinforces the interpretation that the dark energy in these models is not a cosmological constant but a dynamical component. However, at very late times, the strong coupling model exhibits a positive slope for the $Om(z)$ parameter, indicating the phantom field dominance at least at the very late phase of cosmic evolution.

The state finder parameters $(r, s)$ offer a powerful diagnostic tool for distinguishing between different dark energy models by characterizing the geometric evolution of the Universe beyond the standard geometrical parameters such as the Hubble parameter and deceleration parameter \cite{Sahni_2003_77,DEBNATH2025117074}. Defined as

\begin{equation}
   r = \frac{\dddot{a}}{aH^3}, \quad s = \frac{r - 1}{3(q - \frac{1}{2})}, 
\end{equation}
where $q = -\frac{\ddot{a}a}{\dot{a}^2}$ is the deceleration parameter, these quantities form a phase-space known as the statefinder plane. In this plane, the standard $\Lambda$CDM model corresponds to the fixed point $(r = 1, s = 0)$, serving as a benchmark against which alternative cosmologies can be assessed.

\begin{widetext}
    \begin{figure}[H]
\centering
\includegraphics[width=19cm]{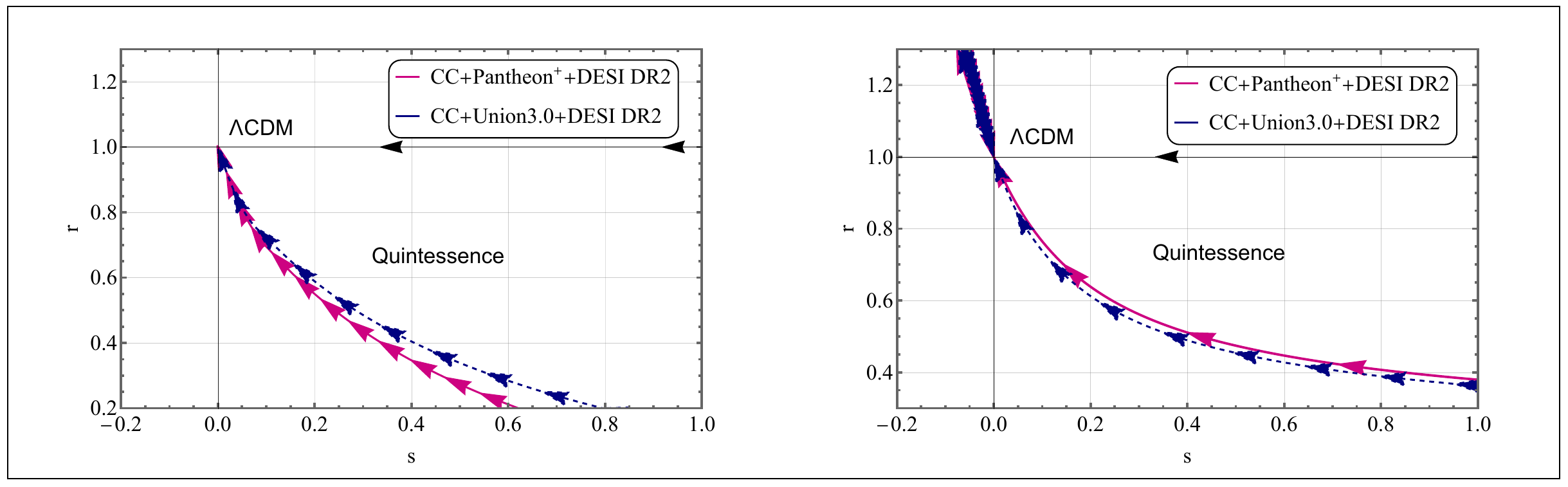}
\caption{{Behavior of statefinder Parameter. Logarithmic: Left Panel, Strong Coupling: Right Panel.}}
\label{Fig7}
\end{figure}
\end{widetext}

In our analysis, the evolution of the statefinder pair $(r, s)$ is plotted in Fig. \ref{Fig7} for both the Logarithmic and the Strong coupling models. The trajectory of the Logarithmic model lies in the region corresponding to quintessence-like behavior ($w > -1$) and gradually evolves toward the $\Lambda$CDM fixed point at late times, indicating that the model mimics standard cosmology asymptotically. On the other hand, the Strong coupling model exhibits a more complex dynamical path: it initially behaves like a quintessence model, at late times transitions through the $\Lambda$CDM phase, briefly approaches the region associated with the Chaplygin gas model, and ultimately returns toward the $\Lambda$CDM point. This oscillatory trajectory reflects the non-linear nature of the model and suggests a richer phenomenology compared to simple quintessence.

The age of the Universe serves as a fundamental parameter in cosmology, providing essential insights into the timeline of cosmic evolution and the formation of large-scale structures. In this study, we estimate the present age of the Universe by numerically integrating the inverse Hubble parameter over redshift, as given in \cite{Vagnozzi_2022_36_27}

\begin{equation}
    t_{U}(z) = \int_0^\infty \frac{dz}{(1+z) H(z)},
\end{equation}
where $H(z)$ encapsulates the redshift-dependent expansion rate of the Universe. Using this approach, we estimate the age of the Universe to be approximately 14.013 Gyr and 13.893 Gyr for the Logarithmic model, and 13.818 Gyr and 13.739 Gyr for the Strong Coupling model, corresponding to the CC+Union3.0+DESI DR2 and CC+Pantheon$^+$+DESI DR2 datasets, respectively. These results are derived using the best-fit parameters from our analysis and are consistent with current observational constraints.

Our results align well with age estimates from other independent probes, including measurements from the Cosmic Microwave Background (CMB) observations, such as those reported by the Planck satellite \cite{Plank_result_2018}. Additionally, these estimates are in agreement with ages derived from the oldest known stellar populations and globular clusters, as well as from radioactive dating of primordial elements.

This consistency across various methodologies reinforces the reliability of our models and their capacity to reproduce realistic cosmic timelines. Accurately determining the age of the Universe is not only vital for validating cosmological models but also for tracing the development of galaxies, stars and other astrophysical structures throughout cosmic history.

\begin{widetext}
    \begin{figure}[H]
\centering
\includegraphics[width=19cm]{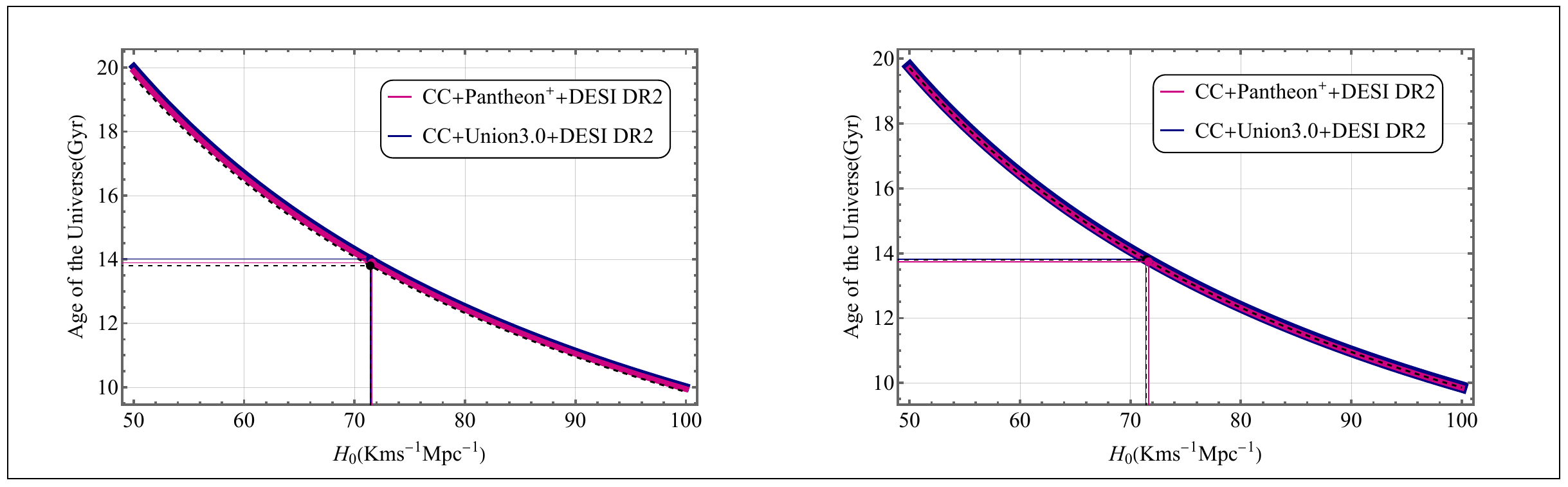}
\caption{{Behavior of Age of the Universe. Logarithmic: Left Panel, Strong Coupling: Right Panel.}}
\label{Fig8}
\end{figure}
\end{widetext}

\begin{widetext}
    \begin{figure}[H]
\centering
\includegraphics[width=18cm]{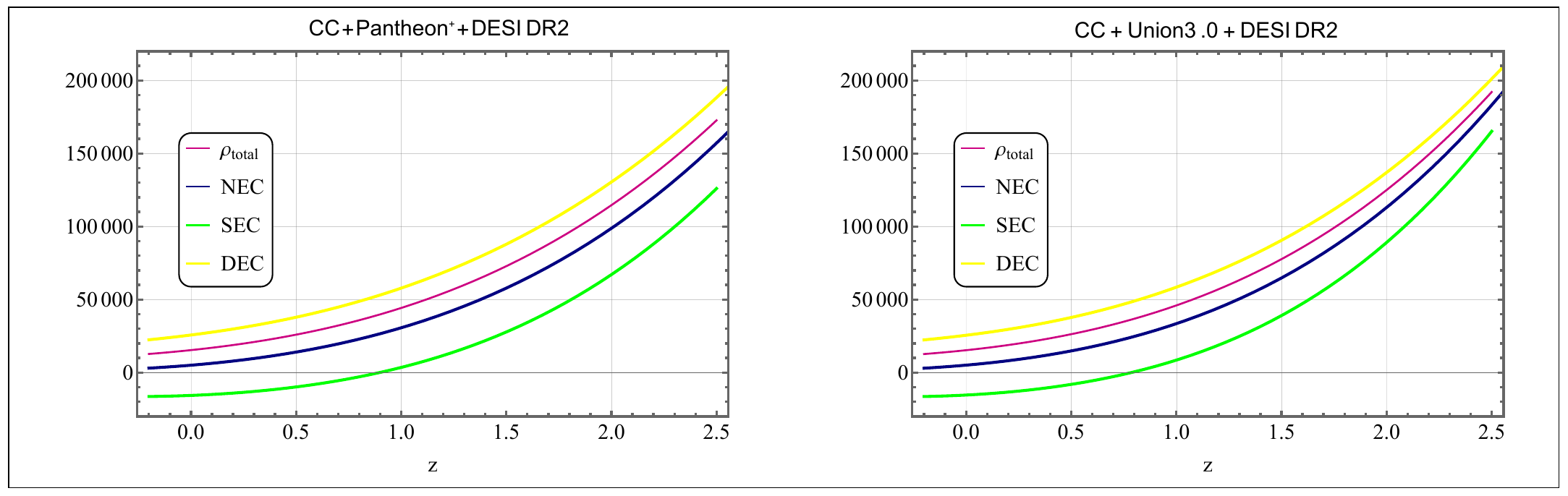}
\includegraphics[width=18cm]{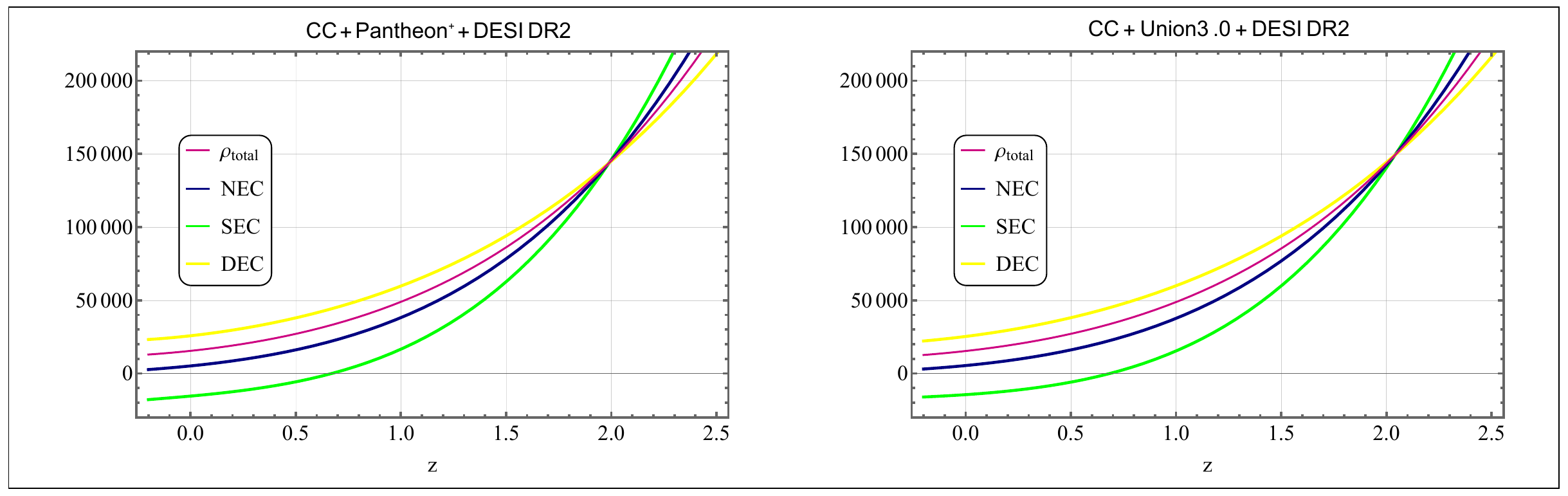}
\caption{{Behavior of Energy Conditions. Logarithmic: Upper Panel, Strong Coupling: Lower Panel.}}
\label{Fig9}
\end{figure}
\end{widetext}

In the context of GR, energy conditions serve as foundational criteria for ensuring physically meaningful cosmological models. These include the Null Energy Condition (NEC), Weak Energy Condition (WEC), Dominant Energy Condition (DEC), and the Strong Energy Condition (SEC) \cite{Carroll_2003_68_023509,Capozziello_2018_781_Erg}. In our analysis, both the logarithmic and Strong coupling models satisfy the NEC, WEC and DEC, which are expressed through inequalities such as $\rho_{\text{tot}} \geq 0$ and $\rho_{\text{total}} \pm p_{\text{total}} \geq 0$, ensuring the non-negativity of energy density and the causal propagation of energy. However, these models violate the SEC, which is given by $\rho_{\text{total}} + 3p_{\text{total}} \geq 0$ (Fig \ref{Fig9}). Another interesting aspect is that, for both the models, SEC is violated after a time frame around which the transition from the deceleration to acceleration occurs. The violation of the SEC is a crucial indicator of cosmic acceleration, as it allows for the emergence of negative effective pressure, necessary to drive the accelerated expansion observed in the current epoch. The violation becomes evident at late times, consistent with the behavior of the deceleration parameter $q(z)$ and the equation of state parameter $\omega$, both of which suggest a transition from decelerated to accelerated expansion. Thus, while maintaining physical consistency through the satisfaction of NEC, WEC and DEC, the controlled violation of the SEC in both models not only enables accelerated cosmic dynamics but also aligns with current observational data \cite{Capozziello_2019_28_1930016}.

\section{Summary and Conclusion}\label{Sec:5} 

In this work, we have presented a comprehensive cosmological investigation of Weyl-type $f(Q, T)$ gravity, a modified gravitational theory that extends the symmetric teleparallel framework by incorporating a coupling between the nonmetricity scalar $Q$ and the trace of the energy-momentum tensor $T$, all within the geometrical setting of Weyl integrable spacetime. We studied two distinct formulations of this theory: a logarithmic model and a Strong coupling model, each characterized by unique dynamical features and coupling parameters that enrich the gravitational action.

We derived the corresponding field equations and reformulated them into a system of first-order differential equations in terms of redshift $z$, enabling a direct comparison with observational datasets. Using the MCMC approach, we constrained the free parameters of each model by performing a joint statistical analysis on the CC+Union3.0+DESI DR2 dataset and the CC+Pantheon$^+$+DESI DR2 measurements. The resulting best-fit parameters were used to numerically reconstruct the cosmological evolution of key observables.

Our findings demonstrate that both models naturally allow for a transition from decelerated to accelerated expansion, consistent with current observational data. The reconstructed Hubble parameter $H(z)$, and all cosmological parameters exhibit behavior in strong agreement with the standard $\Lambda$CDM model at low redshifts. Notably, the Strong coupling model deviates significantly both at low and at high redshift regimes, indicating the possibility of non-standard dynamics during the early stages and very late stage of cosmic evolution. The Logarithmic model, meanwhile, closely tracks $\Lambda$CDM throughout, suggesting its viability as a minimally extended alternative to the cosmological constant.

We examined diagnostic tools such as the equation of state $w(z)$, Om(z), and the statefinder parameters $(r, s)$. These diagnostics further revealed that both models evolve through quintessence-like phases and asymptotically approach the $\Lambda$CDM fixed point at late times. Interestingly, the strong coupling model showed more complex behavior in the statefinder plane, temporarily mimicking Chaplygin gas like features before overlapping to $\Lambda$CDM behaviour, highlighting the dynamical richness of the model.

A particularly important result is the violation of the SEC in both models, while the Null, Weak, and Dominant Energy Conditions remain satisfied. This selective violation provides a theoretical foundation for late-time cosmic acceleration within a purely geometric framework, without resorting to exotic scalar fields or dark energy fluids. The computed age of the Universe, derived from integration over the Hubble parameter, aligns well with observational estimates from Planck CMB data and age-dating of the oldest stellar populations.

Altogether, our results affirm that Weyl-type $f(Q, T)$ gravity offers a viable and physically consistent alternative to the $\Lambda$CDM paradigm. It not only preserves the successes of standard cosmology but also provides room for resolving persistent theoretical challenges, such as the cosmological constant problem. Further, the present analysis within the given set up of the $f(Q, T)$ gravity invites a deeper understanding both at the early and very late phase of cosmic evolution. The geometric flexibility of the formalism and its compatibility with current data make it a robust framework for further exploration.

\section*{Acknowledgement} RB acknowledges the financial support provided by the University Grants Commission (UGC) through Junior Research Fellowship UGC-Ref. No.: 211610028858 to carry out the research work. BM and SKT acknowledge the support of IUCAA, Pune (India) through the visiting associateship program. The authors are thankful to the anonymous reviewers for their comments and suggestions to improve the quality of the paper.

\section*{Appendix-I}\label{Appendix_I}
\begin{figure}[H]
    \centering
    \includegraphics[width=9cm]{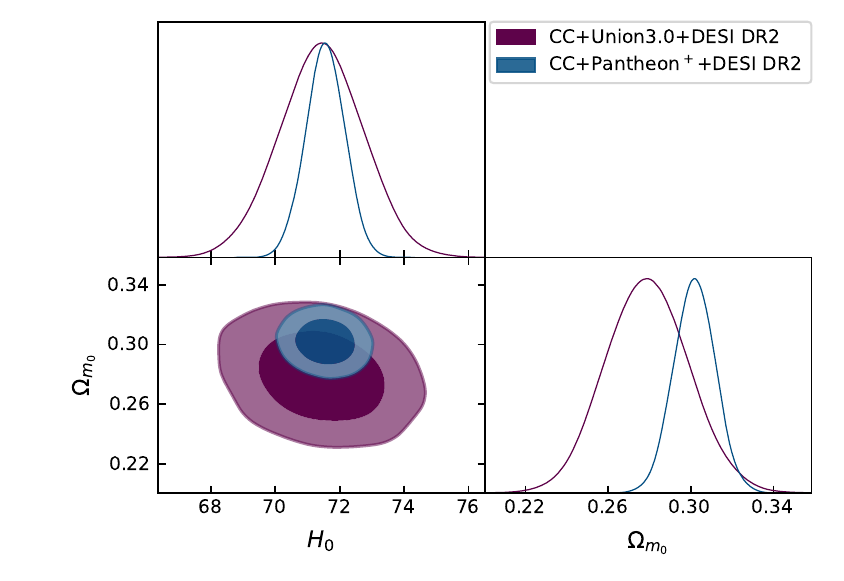}
    \caption{$\Lambda$CDM contour plot}
    \label{Fig10} 
\end{figure}

 In our comparative analysis of the Logarithmic and Strong Coupling Weyl-type $f(Q, T)$ gravity models using the CC+Union3.0+DESI DR2 and CC+Pantheon$^+$+DESI DR2 datasets, we also include the corresponding $\Lambda$CDM results for reference and clarity. Fig.~\ref{Fig10} presents the MCMC posterior distributions and confidence contours obtained under different priors for each dataset, demonstrating good convergence in all cases. From this analysis, we derive the best-fit values of the Hubble constant and matter density parameter as $H_0 = 71.42 \pm 1.3 \, \text{km\,s}^{-1}\text{Mpc}^{-1}$ and $\Omega_{m,0} = 0.277^{+0.021}_{-0.023}$ for the CC+Union3.0+DESI DR2 dataset, and $H_0 = 71.56 \pm 0.62 \, \text{km\,s}^{-1}\text{Mpc}^{-1}$ and $\Omega_{m,0} = 0.308^{+0.019}_{-0.017}$ for the CC+Pantheon$^+$+DESI DR2 dataset. The corresponding present ages of the Universe are estimated as 13.801~Gyr and 13.516~Gyr, respectively. These results suggest that moving toward higher redshifts corresponds to an increase in $H$ and $\Omega_{m}$, accompanied by a decrease in the cosmic age. The $\Lambda$CDM predictions are shown for comparison as the dotted black line in the figure.

\begin{widetext}

 \section*{Appendix-II}\label{Appendix-II}
 \subsection{Model I Log Model}

 \begin{multline}\label{Eq.A1}
  \frac{dH}{dz} = \Bigg( -\beta \varphi^2 + 2 \beta H^2 
    + \frac{-6 \varphi^2 (2 (\beta - 1) \lambda + 7 \beta + 5) 
    - 12 (\beta - 1) H^2 \lambda + 12 H \varphi (2 (\beta - 1) \lambda + 5 \beta + 4)}
    {6 \log \left(\frac{\alpha \varphi_0^2}{\varphi^2}\right) + m^2 - 18} \\
    + 4 H^2 + 2 \beta H \varphi - 2 H \varphi + \varphi^2 \Bigg) \Bigg/ 
    \Bigg( (\beta + 2) H (z + 1) \Bigg)
\end{multline}
     
\begin{multline}\label{Eq.A2}
    \frac{d\varphi}{dz} = \Bigg( -12 (\beta -1) H^2 \lambda 
    - 6 \varphi \left(2 \beta \varphi + (5 \beta + 4) (-H) + \varphi \right) 
    \log \left(\frac{\alpha {\varphi_0}^2}{\varphi ^2}\right)
    + H \varphi \left(24 (\beta -1) \lambda + (5 \beta + 4) 
    \left(m^2 - 6\right)\right) \\
    - \varphi ^2 \left(6 \left(2 (\beta -1) \lambda + \beta + 2\right) 
    + (2 \beta + 1) m^2 \right) \Bigg) \Bigg/ 
    \Bigg( (\beta + 2) H (z+1) \left(6 \log \left(\frac{\alpha {\varphi_0}^2}{\varphi ^2}\right) 
    + m^2 - 18\right) \Bigg)
\end{multline}
       
\begin{equation}\label{Eq.A3}
    \frac{d\lambda}{dz} =  \varphi  \left(6 \log \left(\frac{\alpha  {\varphi_0}^2}{\varphi ^2}\right)+12 \lambda +m^2-6\right) \Bigg/ 6 H (z+1)
\end{equation}

\begin{equation}\label{Eq.A4}
    \rho = 12 H^2 \lambda +6 \varphi  (\varphi -2 \varphi H) \log \left(\frac{\alpha  {\varphi_0}^2}{\varphi ^2}\right)-2 H \varphi  \left(12 \lambda +m^2-6\right)+\varphi ^2 \left(12 \lambda +m^2\right) \Bigg/ \beta +2
\end{equation}

\subsection{Model II: Strong coupling Model}

\begin{multline}\label{Eq.A5}
    \frac{dH}{dz} = \Bigg({H_0}^2 \Big(48 \gamma H^4 \lambda \left(8 \gamma \varphi^2 + {H_0}^2\right) 
    - 8 \gamma H^3 \varphi \left(21 {H_0}^2 \lambda + 2 \gamma \varphi^2 \left(75 \lambda + 2 m^2\right)\right) \\
    + 4 H^2 \left({H_0}^4 \left(3 \lambda + m^2\right) + 3 \gamma {H_0}^2 \varphi^2 \left(22 \lambda + 3 m^2\right) + 30 \gamma^2 \varphi^4 \left(12 \lambda + m^2\right)\right) \\
    - 2 H \varphi \left(12 \lambda + m^2\right) \left(34 \gamma^2 \varphi^4 + {H_0}^4 + 9 \gamma {H_0}^2 \varphi^2\right) 
    + \varphi^2 \left(12 \lambda + m^2\right) \left(16 \gamma^2 \varphi^4 + {H_0}^4 + 6 \gamma {H_0}^2 \varphi^2\right)\Big)\Bigg) \\
    \Bigg/ \Bigg( 2 H (z+1) \Big(2 \gamma {H_0}^4 \left(6 H^2 \lambda + \varphi \left(36 \lambda \varphi - 2 \lambda + 3 m^2 \varphi \right)\right) \\
    + 4 \gamma^2 {H_0}^2 \varphi^2 \left(24 H^2 \lambda - 2 H \left(12 \lambda \varphi + \lambda + m^2 \varphi\right) + \varphi \left(36 \lambda \varphi - 5 \lambda + 3 m^2 \varphi \right)\right) \\
    - 16 \gamma^3 \lambda \varphi^4 (2 H + \varphi) + \varphi{H_0}^6 \left(12 \lambda + m^2\right)\Big)\Bigg)
\end{multline}

\begin{multline}\label{Eq.A6}
    \frac{d\varphi}{dz} = - \Bigg( 16 \gamma  H^3 \lambda  \varphi 
    \left(-8 \gamma ^2 \varphi ^3 + 6 {H_0}^4 + \gamma  {H_0}^2 \varphi  (15 \varphi - 2)\right) \\
    + 4 H^2 \left(32 \gamma ^3 \lambda  \varphi ^5 + 9 {H_0}^6 \lambda 
    + \gamma  {H_0}^4 \varphi  \left(4 \lambda  (3 \varphi - 1) - 3 m^2 \varphi \right) 
    - 2 \gamma ^2 {H_0}^2 \varphi ^3 \left(4 \lambda + 3 \varphi  \left(4 \lambda + m^2\right)\right)\right) \\
    - 2 H \varphi  \left(-16 \gamma ^3 \lambda  \varphi ^5 + 2 {H_0}^6 \left(12 \lambda + m^2\right) 
    + 3 \gamma  {H_0}^4 \varphi  \left(4 \lambda  (9 \varphi - 1) + 3 m^2 \varphi \right) 
    + 2 \gamma ^2 {H_0}^2 \varphi ^3 \left(\lambda  (60 \varphi - 26) + 5 m^2 \varphi \right)\right) \\
    + \varphi ^2 \left(4 \gamma  \varphi ^2 + {H_0}^2\right) 
    \left(-8 \gamma ^2 \lambda  \varphi ^3 + {H_0}^4 \left(12 \lambda + m^2\right) 
    + 2 \gamma  {H_0}^2 \varphi  \left(4 \lambda  (3 \varphi - 1) + m^2 \varphi \right)\right) \Bigg) \\
    \Bigg/ \Bigg( 2 H (z+1) 
    \left(2 \gamma  {H_0}^4 \left(6 H^2 \lambda + \varphi  
    \left(36 \lambda  \varphi - 2 \lambda + 3 m^2 \varphi \right)\right) 
    + 4 \gamma ^2 {H_0}^2 \varphi ^2 \left(24 H^2 \lambda - 2 H 
    \left(12 \lambda  \varphi + \lambda + m^2 \varphi \right) \right. \right. \\
    + \left. \left. \varphi  \left(36 \lambda  \varphi - 5 \lambda + 3 m^2 \varphi \right)\right) 
    - 16 \gamma ^3 \lambda  \varphi ^4 (2 H + \varphi ) + {H_0}^6 \left(12 \lambda + m^2\right)\right) \Bigg)
\end{multline}

\begin{equation}\label{Eq.A7}
    \frac{d\lambda}{dz} = \varphi  \left(2 \gamma  \left(6 H^2 \lambda 
    + \varphi ^2 \left(12 \lambda + m^2\right)\right) 
    + {H_0}^2 \left(12 \lambda + m^2\right)\right) \Bigg/
    {6 H (z+1) \left(\gamma  \varphi  (2 H + \varphi ) +\varphi {H_0}^2\right)}
\end{equation}
    
\begin{equation}\label{Eq.A8}
    \rho = {{H_0}^2 \left(12 H^2 \lambda -2 H \varphi  \left(12 \lambda +m^2\right)+\varphi ^2 \left(12 \lambda +m^2\right)\right)}\Bigg/{2 \left(\gamma  \varphi  (2 H+\varphi )+{H_0}^2\right)}
\end{equation}

\end{widetext}

% \section*{References}
% \bibliographystyle{utphys}
% \bibliography{references}
\section*{References}
%  \bibliographystyle{utphys}
% \bibliography{references}
% \providecommand{\href}[2]{#2}\begingroup\raggedright

%\endgroup

\end{document}